\documentclass[conference]{IEEEtran}
\IEEEoverridecommandlockouts
\usepackage{cite}
\usepackage{amsmath,amssymb,amsfonts}
\usepackage{algorithmic}
\usepackage{graphicx}
\usepackage{textcomp}
\def\BibTeX{{\rm B\kern-.05em{\sc i\kern-.025em b}\kern-.08em
    T\kern-.1667em\lower.7ex\hbox{E}\kern-.125emX}}
\usepackage{multirow}
\usepackage[table, xcdraw, dvipsnames]{xcolor}
\usepackage{booktabs,graphicx,makecell,siunitx,eqparbox}

\usepackage[hidelinks]{hyperref}
\usepackage{chngpage}
\usepackage{geometry}
\usepackage[normalem]{ulem}

\begin{document}

\title{
Enhancement of Anime Imaging Enlargement using Modified Super-Resolution CNN
%Enlargement and Improvement of Anime Images using Modified Super Resolution 
%\\Convolutional Neural Network\\
{\footnotesize}
\thanks{Corresponding author: kuntpong@it.kmitl.ac.th}}

\author{\IEEEauthorblockN{Tanakit Intaniyom\IEEEauthorrefmark{1} \hspace{0.5cm} Warinthorn Thananporn\IEEEauthorrefmark{2} \hspace{0.35cm} Kuntpong Woraratpanya\IEEEauthorrefmark{3}}
\IEEEauthorblockA{Faculty of Information Technology\\
King Mongkut's Institute of Technology Ladkrabang\\ 
Bangkok, Thailand\\
Email: \{imtanakit\IEEEauthorrefmark{1}, t.warinthorn84\IEEEauthorrefmark{2}\}@gmail.com,
kuntpong@it.kmitl.ac.th\IEEEauthorrefmark{3}}
}

% \thanks{Corresponding author: xxxxxxxxxx}}

% \author{\IEEEauthorblockN{xxxxxxxxxx\IEEEauthorrefmark{1} \hspace{0.5cm} xxxxxxxxxx\IEEEauthorrefmark{2} \hspace{0.35cm} xxxxxxxxxx\IEEEauthorrefmark{3}}
% \IEEEauthorblockA{xxxxxxxxxx\\
% xxxxxxxxxx\\ 
% xxxxxxxxxx\\
% Email: \{xxxxxxxxxx\IEEEauthorrefmark{1}, xxxxxxxxxx\IEEEauthorrefmark{2}\}xxxxxxxxxx,
% xxxxxxxxxx\IEEEauthorrefmark{3}}
% }

\maketitle

\begin{abstract}
Anime is a storytelling medium similar to movies and books. Anime images are a kind of artworks, which are almost entirely drawn by hand. Hence, reproducing existing Anime with larger sizes and higher quality images is expensive. Therefore, we proposed a model based on convolutional neural networks to extract outstanding features of images, enlarge those images, and enhance the quality of Anime images. We trained the model with a training set of 160 images and a validation set of 20 images. We tested the trained model with a testing set of 20 images. The experimental results indicated that our model successfully enhanced the image quality with a larger image-size when compared with the common existing image enlargement and the original SRCNN method. 
%Anime is storytelling mediums equal to movies and books. Anime images are a kind of artwork, which is almost entirely drawn by hand. Hence, reproducing existing anime images with larger sizes and higher quality images is expensive. We proposed a model based on convolutional neural networks to extract outstanding features of images, enlarge those images, and enhance the quality of images. We trained the model with 180 images for a training and validation set. 20 images are used for a testing set. The experimental results indicated that our model successfully enlarged and improved image quality better than the common existing image enlargement and original SRCNN methods. 
%Code is available at https://github.com/TanakitInt/SRCNN-anime
\end{abstract}


\begin{figure}[htbp]
\centerline{\includegraphics[width=0.95\linewidth]{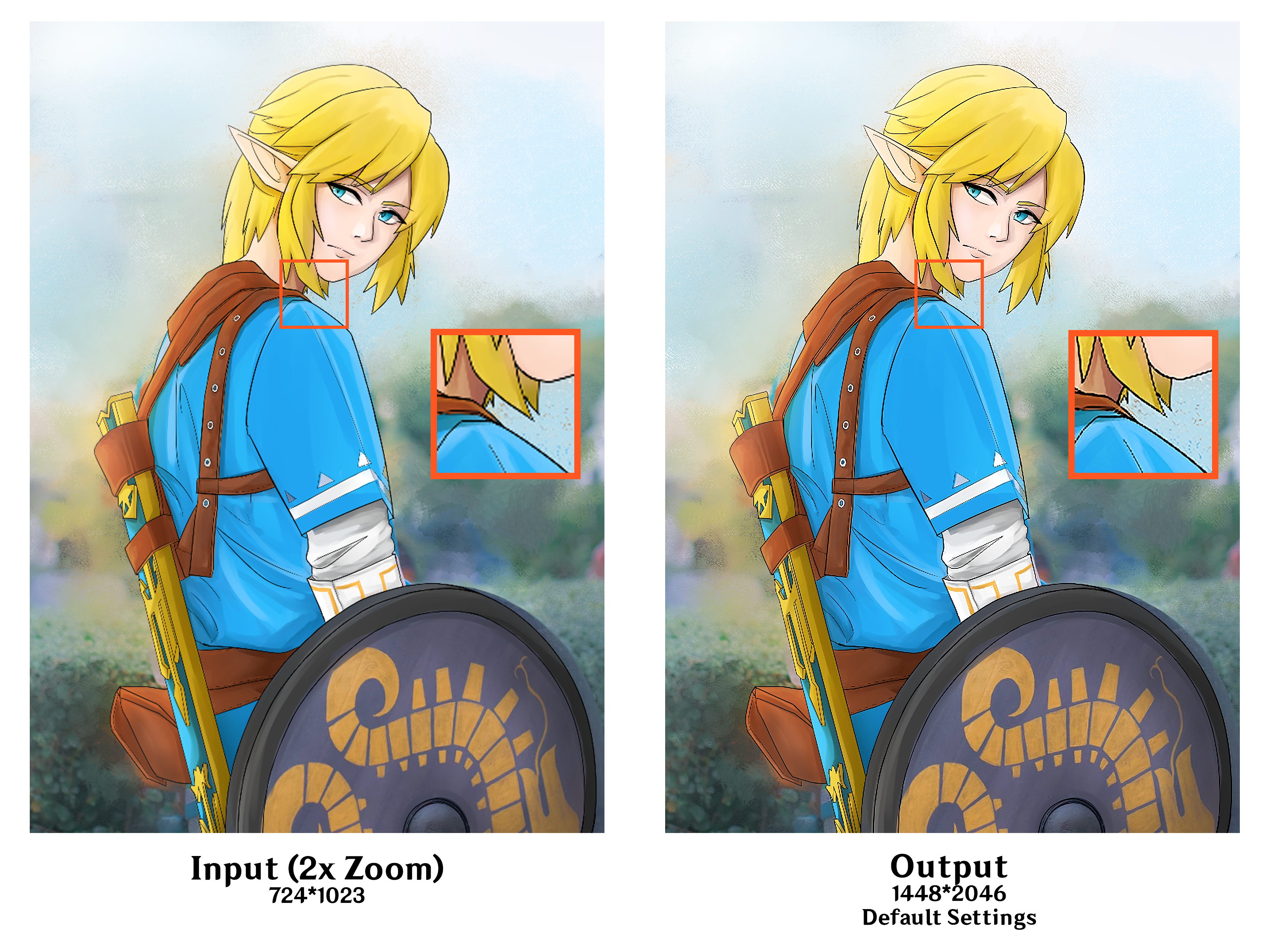}}
\caption{Preview of input and output image in the experiment.}
\label{fig:fig1}
\end{figure}

\section{Introduction}
Art images are a descriptive medium without using text to explain their meaning. Artists create each image or illustration always containing its value and meaning. One of these is Anime-style art images, which are storytelling mediums, like movies and books. Anime is a kind of artwork image, which is almost entirely drawn by hand. It combines graphic art, characterization, cinematography, and other forms of imaginative and individualistic techniques \cite{craig2000japan} along with Manga, a comic typically printed in black-and-white, displayed as a still image, and letters often placed in a format of picture sequences \cite{10.1145/3011549.3011551}. Both Anime and Manga are originated from Japan. In the past, it was impossible to re-create Anime as the same as the original one \cite{liu2021paint}.

In the current digital era, a high resolution for storing and publishing art images is essential. Nevertheless, the former illustrations, currently published in the online media, may have a lower resolution. These images may 
look pixelated or blur. 
Applying existing image enhancement methods on lower resolution images may help fix this issue, but for enlarger images with details, the quality may not be good enough for online media.

%Image is a descriptive medium without using text to explain its meaning. Each image or illustration that the artist created contains its value and meaning. It is impossible to recreate those and match their original. \cite{liu2021paint} %[1]
%An image must have enough resolution for storing information. Older illustrations may not suitably have lower resolution images contained in the online media. These problems may lead to pixelated, blurry, or altered mediums. Using existing image enlargement methods on lower resolution images may help fix those issues but for larger images, image details and quality may not be good enough for online media.

Many of the technical purposes of using low-resolution images are for saving storage, decreasing upload or download speeds from the image sources, deterring illegitimate use of images, and using intentionally them as thumbnails \cite{5957296}. Those purposes of using low-resolution images are not only getting insufficient details in the image itself but also decreasing their values. Fortunately, waifu2x \cite{waifu2x-web}, a free Anime-style art image enlargement and denoising website, and waifu2x program \cite{waifu2x-caffe} for offline use, are available. Fig. \ref{fig:fig1} shows an example of the output enlarged image with high quality. It inspired us to investigate how to improve the quality of enlarged images.

By applying the deep learning model, called Super-Resolution Convolutional Neural Network (SRCNN), as the waifu2x used, the Anime image was enlarged and enhanced its output quality. As our preliminary experiments with waifu2x, 
the output image quality after enlargement was not adequate sometimes. 
%still suffered from the enhancement quality; that is, the  
We have assumptions that this phenomenon happened due to the SRCNN architecture that uses only one enhancement module. 
As mentioned so far, we proposed a modified Super-Resolution CNN (m-SRCNN). This model can enhance the quality of art images and support image enlargement tasks. To achieve our goal (See details in Fig. \ref{fig:fig5}), we provided pre-processing phase to enhance the quality of datasets for training a model; therefore, the model, trained by the high quality of datasets, usually resulted in a high quality output image. Furthermore, we provided four post-processing modules, called image-enhancement only, enlargement and enhancement, double enhancement, and double enlargement, for various purposes.

%\textcolor{blue}{
%As mentioned so far, we proposed a modified Super-Resolution CNN (m-SRCNN). This model can enhance the quality of art images, enlarge with double resolutions, and the enhanced-only task that does not required enlargement. To achieve our goal, we added convolution transpose layers into the SRCNN model, changed filter settings, adjusted kernel sizes of the CNN, and applied Up-Bilinear and Up-Bicubic models individually.}
%As mentioned so far, we proposed a Modified Super-Resolution CNN (m-SRCNN). This model can enhance the quality of art images, enlarged with double resolutions, and the enhanced-only task that does not required enlargement. To achieve our goal, we add convolution transpose layers into the SRCNN model, change filters setting, adjust kernels size of the CNN, and applying Up-Bilinear and Up-Bicubic \textcolor{green}{model} individually.}  
Finally, we tested our methods with Nico-illust dataset \cite{10.1145/3005358.3005388} and evaluated the performance by using the PSNR and SSIM metrics. Results show that our models can outperform the conventional SRCNN and other baseline methods in most cases.
%For the experiment, we test our data with a test set from \cite{10.1145/3005358.3005388} and evaluate the performance by using the PSNR and SSIM indicators. Results show \textcolor{green}{\sout{that}} our models can outperform SRCNN and other baseline methods in most cases.}

%\textcolor{blue}{
%As mentioned so far, we proposed a modified super-resolution CNN for enhancing \textcolor{red}{the} quality of art images, enlarged with double resolutions\textcolor{red}{, and the enhanced-only task that do not required enlargement. To achieve our goal, we add convolution transpose layers into the SRCNN model and change filters and kernel size compared to the original super-resolution CNN.}}  
%The experiments are conducted using existing image enlargement, convolutional neural networks (CNN), and deep learning to enlarge and improve images keeping their quality and details. Our experiments are mainly focused on anime-style art images which are storytelling mediums equal to movies and books. Anime is a kind of artwork, which is almost entirely drawn by hand. It combines graphic art, characterization, cinematography, and other forms of imaginative and individualistic techniques. \cite{craig2000japan} along with manga, a comic typically printed in black-and-white displayed as a still image and letters often place in a format of picture sequence. \cite{10.1145/3011549.3011551} Both anime and manga are originating from Japan.

\section{Related Works}

%\subsection{Digital Image Processing}
%In one kind of signal processing, the image will be digitally converted to numbers called matrix, every single matrix called pixel representing its data. Digital image processing is a mathematical process working on an image to obtain features or detail. To accomplish this, there are matrices called filters which are often 2-dimensional 3 by 3 height and width matrices, doing a mathematics calculation to match the proposal. A method stated above is called convolution. \cite{10.5555/281875} (1) The formula is shown below.

%\begin{equation}
%(f \ast g)(t):=\int_{-\infty}^{\infty} f(\tau) g(t-\tau) d \tau \label{eq:convolution}
%\end{equation}

%Convolutions require a filter matrix by convolution filter and image matrix. Depending on filters, multiple features can be extracted, for example – sharpening, blurring, and denoising. Image processing can be applied to detection systems, image quality improvement, image segmentation, or other related fields of study. \cite{shapiro2001computer}

\subsection{Image Scaling and Interpolation}
%Image Scaling}
Image scaling is a geometric transformation. By performing this in image processing, images can be scaled up or down by an image scaling function. An important mechanism behind the image scaling function is an interpolation. Here, we used three common interpolation methods---Nearest Neighbor, Bilinear, and Bicubic \cite{10.1145/2499788.2499859} in our experiments. A nearest neighbor interpolation replaces every pixel with the nearest pixel in the output image. A bilinear method interpolates pixel values by estimating new pixel values in between two pixels. A bicubic method interpolates pixel values as the same as the bilinear method, but it uses third-degree polynomials instead \cite{1163711}. By image scaling up, the pixel amount of the image is increasing, thus resulting in more details. On the other hand, the image scaling down causes the details of the image decreasing.

%Scaling is a subset of geometric transformation. By performing this in image processing, images can be scaled up or down by a scale factor. An important mechanism behind scaling function is interpolation. Here, we used three common interpolation methods-–-Nearest Neighbor, Bilinear, and Bicubic \cite{10.1145/2499788.2499859} in our experiments. A nearest neighbor interpolation replaces every pixel with the nearest pixel in the output, multiple pixels of the same color will present. Bilinear scaling will interpolate pixel values by estimating new pixel values between each pixel. Bicubic scaling will interpolate pixel value as same as the bilinear method, but it uses third-degree polynomials instead. \cite{1163711} By scaling up the image, the pixel amount of the image will increase, resulting in more details, on the other hand, scaling down an image will cause the details of the image to decrease.

\subsection{Super-Resolution (SR)}
A super resolution (SR) is a process that transforms a low-resolution (LR) image into a high-resolution (HR) image. As reported in \cite{7115171}, SR can be categorized into four types---prediction models, edge-based methods, image statistical methods, and patch-based (or example-based) methods. Among them, the patch-based achieved the best performance. Inside this method, the Nearest Neighbor algorithm was used in the low-resolution space for reconstruction. Recently, the researchers have proposed a new method that applied deep learning like CNN instead of the Nearest Neighbor approach, and it can achieve better performance in the image enlargement \cite{7780551}. As reasons mentioned previously, we have chosen the SR algorithm implemented on Convolutional Neural Network for our image enlargement method.

\subsection{SRCNN}
SRCNN is short for a Super-Resolution Convolutional Neural Network. SRCNN is a CNN model built to enhance image quality by learning the end-to-end mapping between bicubic interpolation of LR and HR images. The overall structure consists of three parts: (i) patch extraction and representation, which extracts patches from the input represents as a high-dimensional feature vector, (ii) non-linear mapping, which maps the HR feature, and (iii) reconstruction, which builds up features to form the final output image \cite{7115171}. This designed structure is based on sparse coding, which aims in finding a sparse representation of input data in a form of a linear combination of basic elements \cite{Sparse}.
%These basic elements are called atoms, combined into a dictionary \cite{Sparse}.
%Short for Super-Resolution Convolutional Neural Network. SRCNN is a CNN model built to enhance image quality by learning the end-to-end mapping between bicubic interpolation of LR and HR images. The overall structure consists of three parts based on sparse coding, which aims at finding a sparse representation of the input data in the form of a linear combination of basic elements. Basic elements are called atoms, combined into a dictionary \cite{4587647}. Three parts are the patch extraction and representation part which extracts patches from the input represents as a high-dimensional feature vector, the non-linear mapping part which maps the HR feature, and the reconstruction part which buildup features to form the final output image \cite{7115171}.

\section{Methodology}
\begin{figure}[htbp]
\centerline{\includegraphics[width=0.8\linewidth]{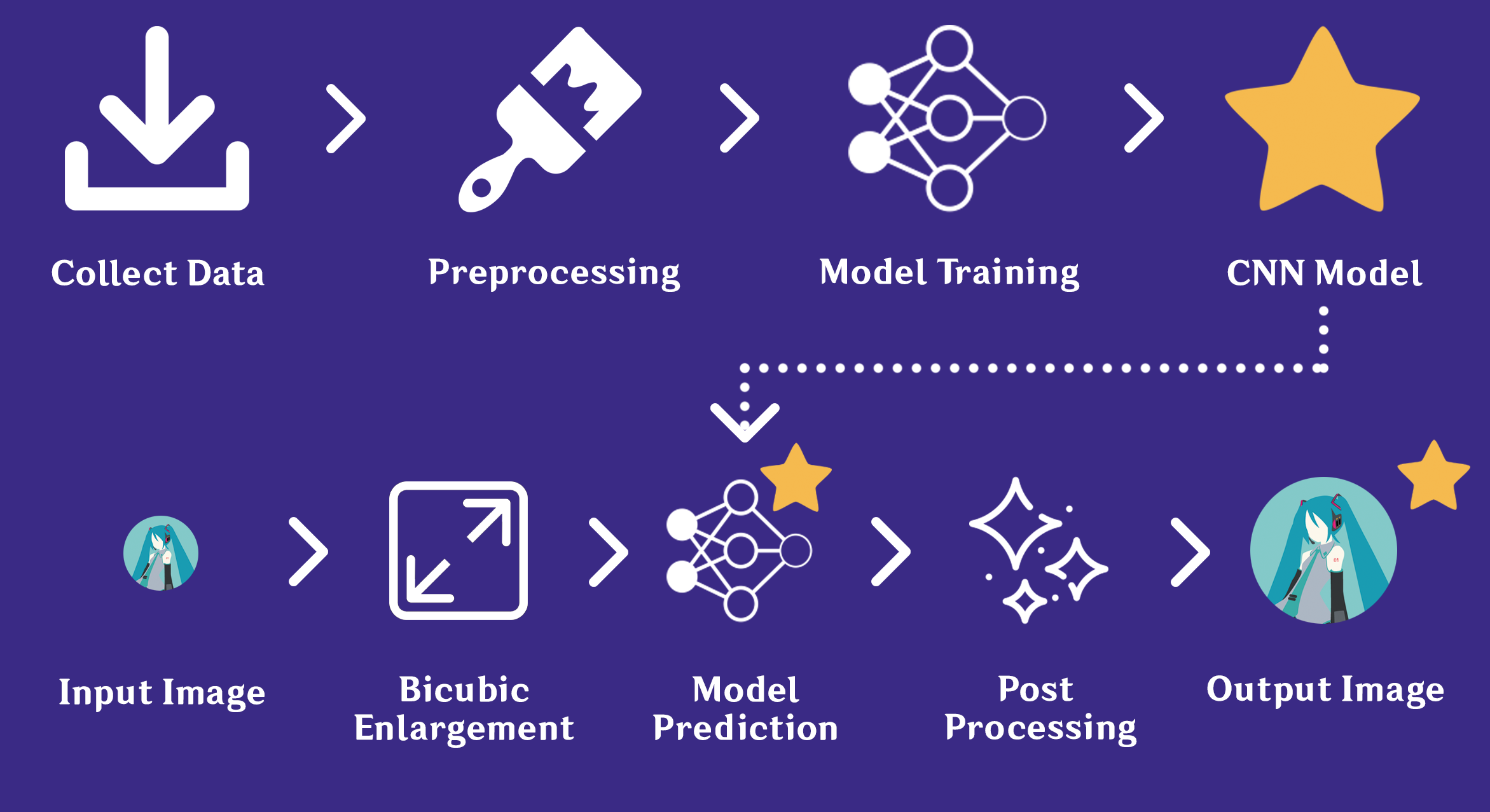}}
\caption{A simplified diagram of our modified super-resolution CNN.}
\label{fig:fig2}
\end{figure}

As we know, basic image upscaling algorithms, such as Nearest Neighbor, Bilinear, and Bicubic, do not achieve in enhancing the quality of enlarged images. An alternative way is using a neural network for image super-resolution, since it can learn to fill details, based on the information collecting from a large set of images. Therefore, here, we proposed a modified super-resolution CNN (m-SRCNN) as shown in Fig.~\ref{fig:fig2} and the followings describe its functional operations and settings. 

\subsection{Data Collection} \label{sub:datacollection}
\begin{figure}[htbp]
\centerline{\includegraphics[width=0.8\linewidth]{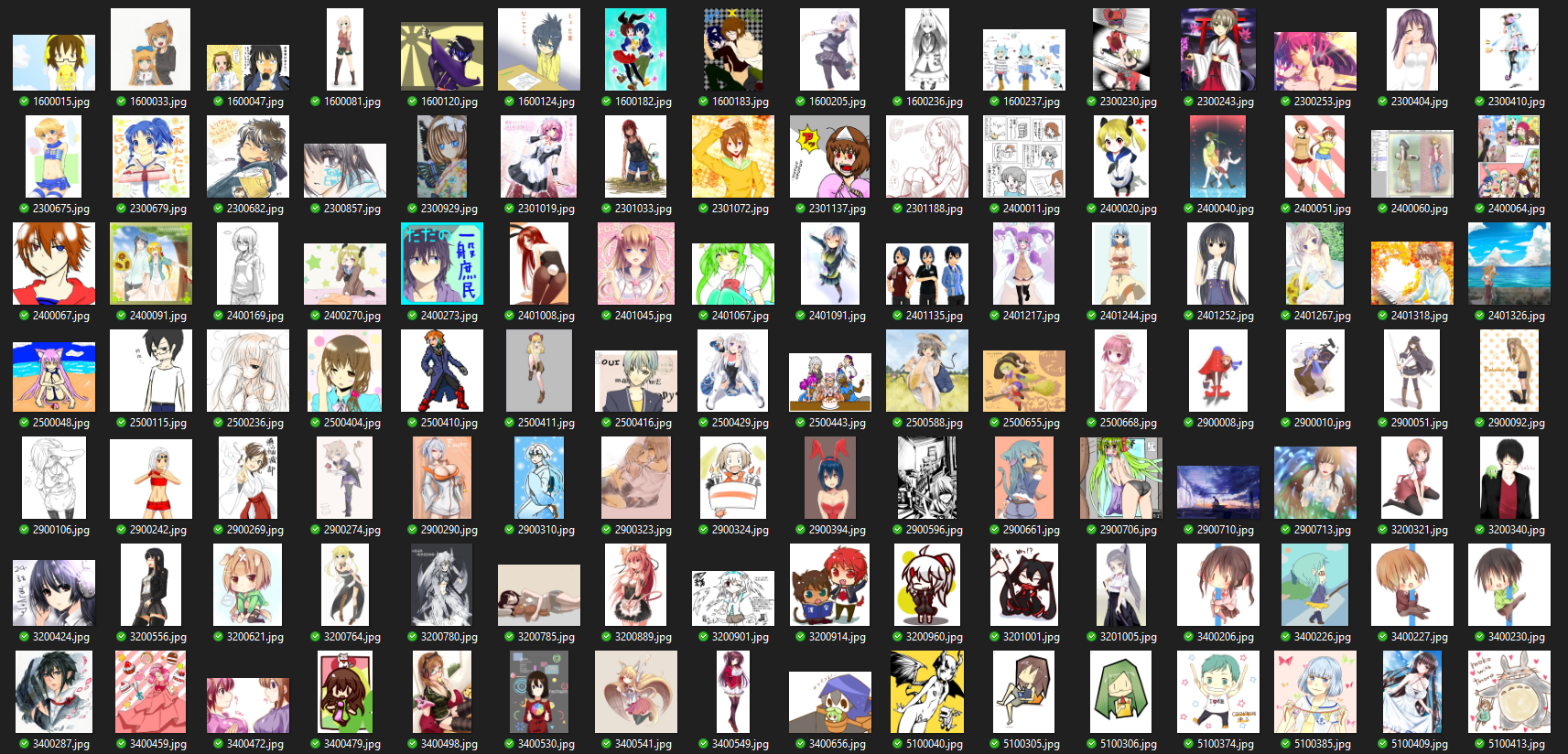}}
\caption{A part of collected images from Nico-illust dataset.}
\label{fig:fig3}
\end{figure}

In the first step, we collected Anime art-style images from Nico-illust \cite{10.1145/3005358.3005388}, a community illustration from Niconico Seiga and Niconico Shunga, for training, validating, and testing sets. The collected images, which are a JPG format, are split into 160 images for a training set, 20 images for validation, and 20 images for a testing set. Fig.~\ref{fig:fig3} shows part of the collected images.
%In the first part, we collect anime art-style images which are for training, validating, and testing sets. Generally, anime art-style images can be various and diverse, but due to hardware limitations, the images in datasets we collect and select will only have some of the whole full datasets.

%The Images we selected from full datasets contain JPG format splitting into a training set for 160 images, validation for 20 images, and testing set for 20 images. Each image will have approximately 600 by 600 pixels varies by each image. All Training, Validating, and Testing set are Nico-illust \cite{10.1145/3005358.3005388}, a community illustration from Niconico Seiga and Niconico Shunga.

\subsection{Image Preprocessing} \label{sub:image-preprocessing}
All collected images were sharpened and denoised to improve their image quality and then they were split into LR and HR. Here, we provided two sets of LR images. The first set were downscaled by a bilinear interpolation with 0.5x, and then they were upscaled by a bilinear interpolation with 2x. Likewise, the second set of LR images was processed in the same way of the first set, but using a bicubic interpolation. 
On the other hand, HR images were original images. All LR and HR images were cropped with the size of 32-by-32 pixels and then were converted to YCbCr color space. Since SR algorithms are only applied on the Y channel, while the Cb and Cr channels are upscaled by bicubic interpolation \cite{7115171}. Lastly, images were stored in h5 file format.
%Before model training, all images in the training and validating set will be passing the image sharpening and denoising process to increase its quality and details. After completing the previous process, all images will be splitting into LR and HR. LR images are downscaled 0.5x by bilinear. then upscaled 2x by bilinear and bicubic interpolation separately. In short, LR images are upsampled to 2x scale coarse HR images. HR images are original images. All LR and HR images will be cropped to 32 by 32 pixels and converted to YCbCr color space. Since SR algorithms are only applied on the Y channel, while the Cb and Cr channels are upscaled by bicubic interpolation. \cite{7115171} Lastly, images are stored in h5 file format.

\subsection{Model Training} \label{sub:model-training}
At this stage, we set up hardware and software for training a model as follows. The hardware specifications were Intel Core i5-11400F CPU, 1 unit of Nvidia GeForce GTX 750 Ti GPU, 16 GB of RAM, and 480 GB of SSD storage, and we used Keras and TensorFlow 2.2 as a deep learning development framework and Python 3.7 as software development, including with CUDA (version 10.1.243) and cuDNN (version 7.6.5), a GPU accelerated functionality, and library for deep neural nets. Original and modified super-resolution CNN configurations can be seen in Tables~\ref{table1} and \ref{table2}, respectively.
%To start a deep learning model training, the first step to do is hardware and software setup. We are using Intel Core i5-11400F CPU, 1 unit of Nvidia GeForce GTX 750 Ti GPU, 16 GB of RAM, and 480 GB of SSD storage. For software, we are using Keras and TensorFlow 2.2 as a deep learning development framework and Python 3.7 for software development including with CUDA (version 10.1.243) and cuDNN (version 7.6.5), a GPU accelerated functionality and library for deep neural nets. Original and modified super-resolution CNN architecture and configurations can be shown below.

\begin{table}[ht!]
\centering
\caption{A configuration of the original SRCNN}
\begin{center}
\begin{tabular}{c c c c c}
\hline
\textbf{Layers} & \textbf{Filters} & \textbf{Kernel} & \textbf{Activation} & \textbf{Bias}      \\
\textbf{}       & \textbf{}        & \textbf{Size}   & \textbf{}           & \textbf{}          \\ \hline
Convolution     & 128              & 9*9            & LeakyReLU           & True               \\
Convolution     & 64               & 3*3             & LeakyReLU           & True               \\
Convolution     & 1                & 5*5             & Sigmoid             & True               \\ \hline
\end{tabular}
\label{table1}
\end{center}
\end{table}

\begin{table}[ht!]
\centering
\caption{A configuration of our m-SRCNN}
\begin{center}
\begin{tabular}{c c c c c}
\hline
\textbf{Layers}       & \textbf{Filters} & \textbf{Kernel} & \textbf{Activation} & \textbf{Bias} \\
\textbf{}             & \textbf{}        & \textbf{Size}   & \textbf{}           & \textbf{}     \\ \hline
Convolution           & 64               & 5*5             & LeakyReLU           & True          \\
Convolution           & 64               & 5*5             & LeakyReLU           & True          \\
Convolution           & 16               & 3*3             & LeakyReLU           & True          \\
Convolution Transpose & 32               & 3*3             & LeakyReLU           & True          \\
Convolution Transpose & 32               & 3*3             & LeakyReLU           & True          \\
Convolution           & 3                & 1*1             & Sigmoid             & True          \\ \hline
\end{tabular}
\label{table2}
\end{center}
\end{table}

\begin{figure*}[ht!]
\centering
\centerline{\includegraphics[width=0.7\linewidth]{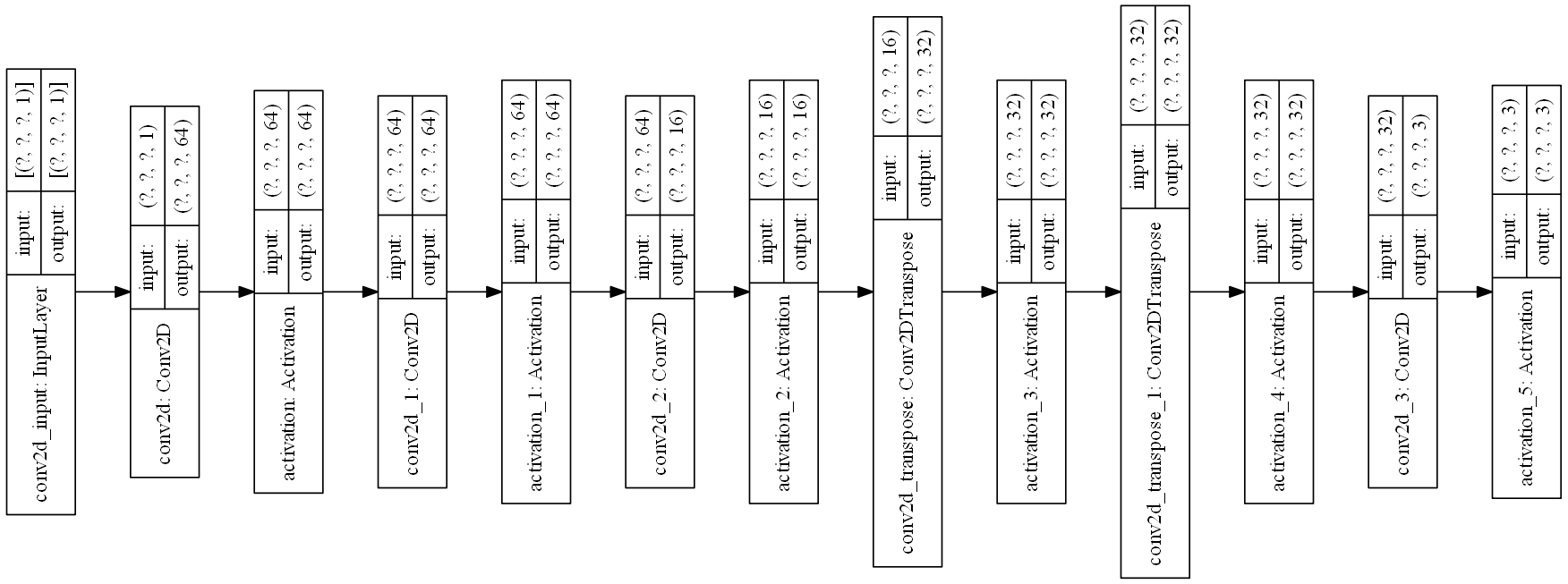}}
\caption{A modified super-resolution CNN architecture.}
\label{fig:fig4}
\end{figure*}

Layers of the original SRCNN were from \cite{7115171} and those of m-SRCNN were from \cite{img-ai}, as visualized in Fig.~\ref{fig:fig4}. The m-SRCNN model was modified and customized by adding the convolution transpose layer and removing the upscaling layer from the deep learning model, when compared to the original SRCNN. In model learning, parameters were set as follows: 50 epochs, 0.003 of the learning rate for m-SRCNN, 32 of batch size. 
%Original SRCNN layers have been reference from referenced research \cite{7115171}. Modified super-resolution CNN layers have been referenced from reference GitHub Repository \cite{img-ai}, the model has been modified and customized by adding convolution transpose compared to the original SRCNN and remove the upscaling layer in the deep learning model compared from reference GitHub Repository. As previously mentioned, our modified super-resolution CNN will only enhance images. During the development, Training using our hardware and software setup, learning parameters are 50 epochs, 0.003 of the learning rate for modified super-resolution CNN, 0.0003 of the learning rate for original super-resolution CNN, 32 of batch size. It took 2 hours to complete for each model training. For optimal training performance, the learning rate for both models has adjusted to different values.

\subsection{Post Processing} \label{sub:post-processing}
At the last stage, output images, from a deep learning model, can be customized by post-processing, especially the image denoising process, developed by using Python 3 OpenCV fastNlMeans \cite{4106808} and Bilateral filter \cite{710815}. Also, image enhancement-only, Double Enhancement, and Double Enlargement functions \cite{iqa-srcnn} are added for image quality improvement. Note that these functions are optional, depending on tasks and user satisfactions. Users may choose to apply these methods or maintain the original output.
%\textcolor{orange}{These function can improve image visual quality, also additional image enhancement function available for user as satisfaction, not limited to the default one.}
%Images from deep learning models can be customized by post-processing processes. Our development has focused on denoising images, which enhance detail in each individual pixel developed using Python 3 OpenCV fastNlMeans \cite{4106808} and Bilateral filter. \cite{710815} Also, the double enhancement and double enlargement function.

\begin{figure*}[htbp]
\centerline{\includegraphics[width=0.7\linewidth]{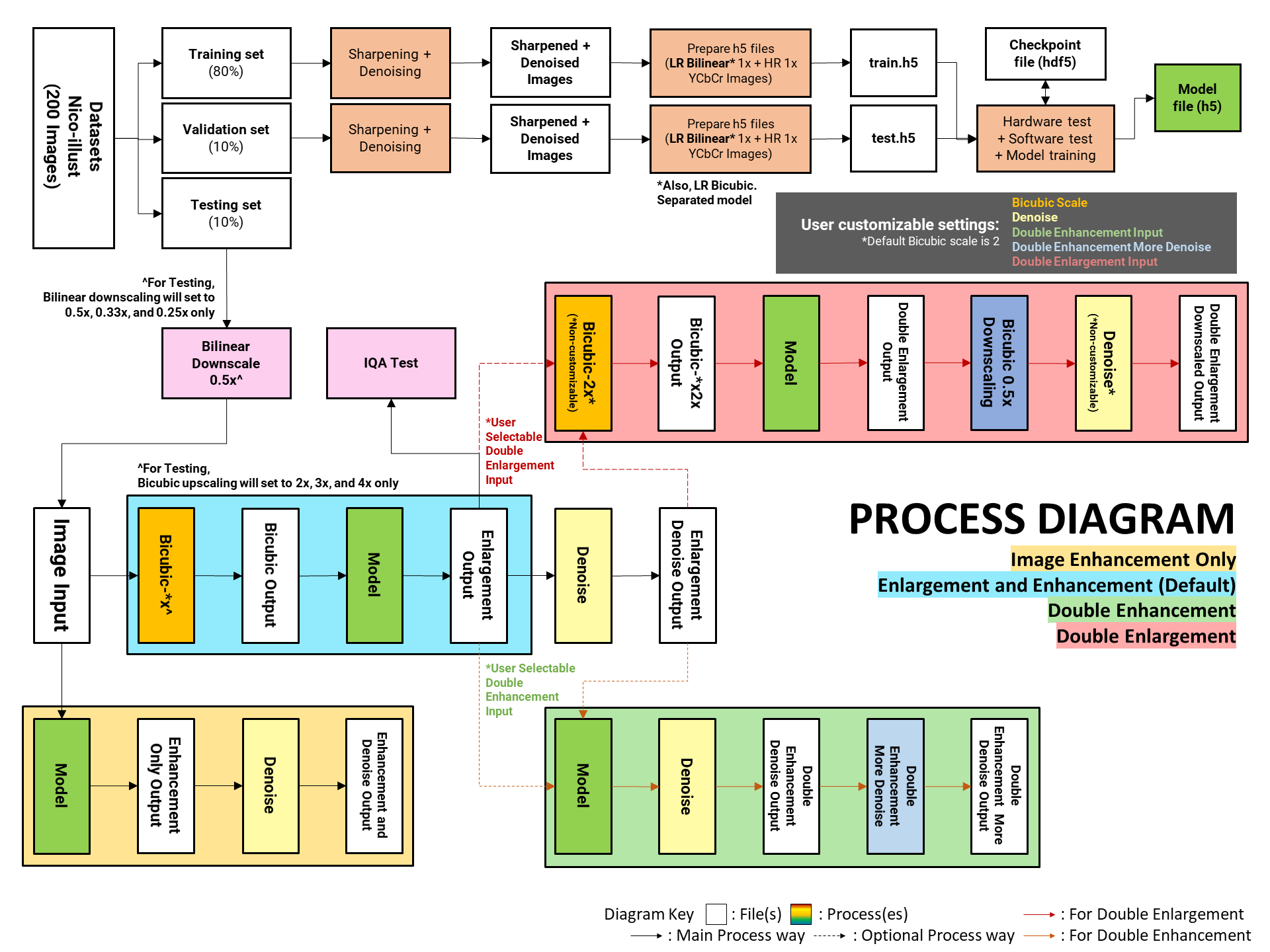}}
\caption{A diagram of all processes of cooperation of m-SRCNN.}
\label{fig:fig5}
\end{figure*}

\section{Experiments} \label{section:experiment}
To evaluate the performance of our method and baselines, including (i) conventional interpolation methods and (ii) an online application like waifu2x \cite{waifu2x-caffe}, two experiments were set up to test each of these aspects as the followings.

\subsection{Data Preparation and Baseline Selection} \label{sub:data-preparation}
We used Nico-illust dataset \cite{10.1145/3005358.3005388} for our experiments. We set the original images with their original size as reference (or ground truth). The input images (or test images) were downscaled by a bilinear interpolation to three different sizes of 0.50x, 0.33x, and 0.25x.

For performance comparison, our baselines are %include 
%the downscaled images were enlarged with our baseline methods, including 
Nearest Neighbor, Bilinear, Bicubic, SRCNN with the up-bilinear model, and SRCNN with the up-bicubic model.
%, modified SRCNN with \textcolor{red}{the} up-bilinear model, and modified SRCNN with \textcolor{red}{the} up-bicubic model. 
Three sizes for image enlargement in testing are 2x, 3x, and 4x upscalings.
%To compare the results and performance of a deep learning model, we must use appropriate methods to benchmarking these images. The benchmark will use the testing set as an independent variable compared with other image enlargement methods by downscaling them to 0.5x, 0.33x. and 0.25x by bilinear interpolation. The original images before downscaling are called Ground Truth. After downscaling, the image will be enlarged with different methods – Nearest Neighbor, Bilinear, Bicubic, SRCNN (up-bilinear model), SRCNN (up-bicubic model), Modified SRCNN (up-bilinear model), and Modified SRCNN (up-bicubic model) with an enlargement scale of 2, 3, and 4 times.

\subsection{Evaluation Metrics} \label{sub:evaluation-metrics}
In model training, we use Mean Squared Error ($MSE$), defined in (1), as a loss function, and use Peak Signal to Noise Ratio ($PSNR$) \cite{10.1007/978-3-319-10593-2_25}, denoted in (2), and Structural Similarity Index Measure ($SSIM$) \cite{1284395}, expressed in (3), as experimental assessment metrics.

\begin{equation}
MSE=\dfrac{1}{mn}\sum ^{m-1}_{i=0}\sum ^{n-1}_{j=0}\left[ I\left( i,j\right) -K\left( i,j\right) \right] ^{2} \label{eq:mse}
\end{equation}
where %$MSE$ is the mean squared error score of the compared images, 
$m$ and $n$ are the size of an image, $i$ and $j$ denote the pixel locations of an image in the Cartesian coordination system, and $I$ and $K$ are the original and enlarged images, respectively.

\begin{equation}
PSNR=20\cdot \log _{10}\left( \dfrac{255}{\sqrt{MSE}}\right) \label{eq:psnr}
\end{equation}

%\textcolor{red}{where $PSNR$ is the peak signal-to-noise ratio quality score of the image, $MAX$ is the maximum possible pixel value of the image, and $MSE$ is the mean squared error from (\ref{eq:mse}).}

\begin{equation}
SSIM\left( x,y\right) =\dfrac{\left( 2\mu _{x}\mu _{y}+c_{1}\right) \left( 2\sigma _{xy}+c_{2}\right) }{\left( \mu _{x}^{2}+\mu _{y}^{2}+c_{1}\right) \left( \sigma _{x}^{2}+\sigma _{y}^{2}+c_{2}\right) } \label{eq:ssim}
\end{equation}
where $x$ and $y$ are two compared images, $\mu$ indicates the average intensity value of all pixels, and $\sigma$ denotes the contrast value, calculated by a standard deviation of all pixels.
%where $SSIM$ is the structural similarity index score of the image, 
%$x$ and $y$ are two compared images, and $\mu$ indicates the luminance that average overall pixel values, $\sigma$ denotes the contrast that calculated by applying standard deviation to all pixel values.}

%\textcolor{red}{\subsection{Experimental Framework and Setting}}
%\textcolor{red}{Fig.~\ref{fig:fig5} show a diagram of all processes of cooperation of m-SRCNN. Image from the user will be enlarged with bicubic interpolation with the desired scale then the deep learning model we trained will enhance it. From this point, users can customize their image getting from the deep learning model. Default settings would be used for image benchmarking. These settings are:}

%\textcolor{red}{
%\begin{itemize}
%\item Bicubic enlargement scale equal 2
%\item No denoising
%\item No image customization or post processing
%\end{itemize}}

\subsection{Experimental Framework and Settings} \label{sub:framework} 
\begin{figure}[htbp]
\centerline{\includegraphics[width=0.6\linewidth]{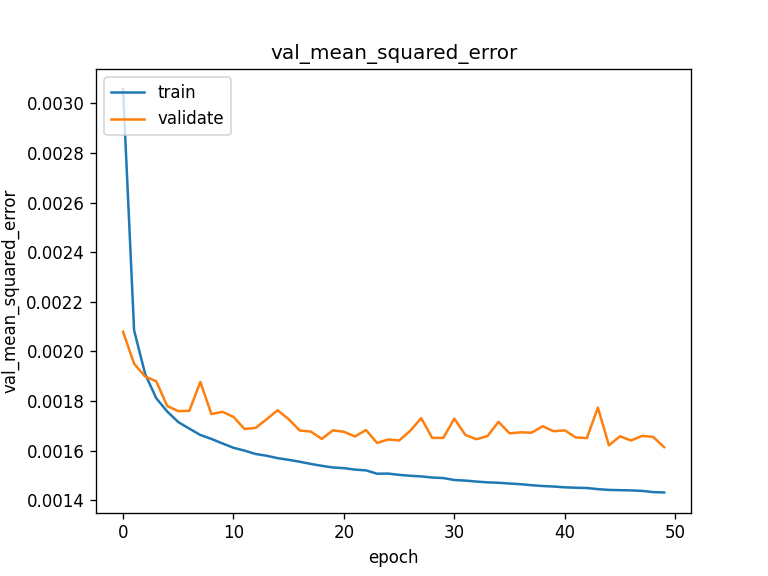}}
\caption{A MSE graph of the m-SRCNN with a bilinear upscaled model.}
\label{fig:fig6}
\end{figure}

%Starting from users input an image and start program to the users get a result. 
Fig.~\ref{fig:fig5} shows a diagram of all processes of cooperation of m-SRCNN.
The followings brief a flow of the experimental diagram. The training and validation image sets were preprocessed as described in~\ref{sub:datacollection} and~\ref{sub:image-preprocessing}. Then all processed images were converted to h5 file format as mentioned in~\ref{sub:image-preprocessing}. The trained model, proposed in~\ref{sub:model-training}, was stored in h5 format and used for testing. 
Fig.~\ref{fig:fig6} shows an example of model training in terms of a MSE graph. All model checkpoints also made and stored as hdf5 format if there were problems occurred during a model training. A testing set of the prepared dataset were bilinear-downscaled as the same as~\ref{sub:data-preparation} to use as an input image in model testing. Next, pre-upsampling image was enlarged by the bicubic method, and then converted from RGB to YCbCr color space for a model prediction. Lastly, the newly saved enhanced image was stored and kept for model evaluation.
%First, training and validation set would be preprocessed as in~\ref{sub:datacollection} and~\ref{sub:image-preprocessing} step then, convert them to h5 file format as mention in~\ref{sub:image-preprocessing} step. The trained model from~\ref{sub:model-training}, stored in h5 format, would be kept for testing. A MSE graph of model training were shown in Fig.~\ref{fig:fig6} as a sample. Model checkpoints also made and stored as hdf5 format if there were problem occurred during a model training. A testing set from prepared dataset would bilinear-downscaled as same as~\ref{sub:data-preparation} step to be used as input image in model testing. Next step, image pre-upsampling, would enlarge input image by bicubic then convert from RGB to YCbCr color space continue to the model prediction. Lastly, newly saved enhanced image was stored and kept for model evaluation.}

%\textcolor{red}{For testing as seen in Fig.~\ref{fig:fig5}, test image from Nico-illust Datasets\cite{10.1145/3005358.3005388} will be bilinear downscale with 0.5x, 0.33x, and 0.25x scale then bicubic upscale with 2x, 3x, and 4x scale as same as baseline methods. After upscaling, apply trained model as seen in Fig.~\ref{fig:fig6}, configured with Table.~\ref{table2} to the image.}
%\textcolor{orange}{\sout{For benchmarking, image will be bicubic-interpolated with scale of 2 into the model. Then, the model will enhance the image. For a model testing, no furture image denoising, image customization, or other post processing has been made in the benchmarking.}}

\subsection{Results}
%By comparing MSE in different epochs, we can conclude that each epoch of model training is continuously improving.

%\subsection{Denoising}
%An image post-processing function, optional and customizable function for users. Generally, results might improve image quality, but results may vary. The denoising option can be set in image selection for both double enhancement and double enlargement functions.

%\subsection{Additional features and Unseen illustrations without Ground Truth testing}
%A customizable post-processing option is available for users. Generally, results might improve image quality, but results may vary. These are Double enhancement and Double enlargement functions.

In the first experiment, as can be seen in Table~\ref{table3}, at 2x upscaling, our m-SRCNN in both models, up-bilinear and up-bicubic, outperform all test baselines and the original SRCNN, especially m-SRCNN with up-bilinear model. Nonetheless, at 3x upscaling, the original SRCNN with up-bicubic model is slightly better than our method, m-SRCNN with up-bicubic. When regarding in overall performance of all test methods, the average PSNR and SSIM are much reduced. As expected, when compared with 3x upscaling, the overall performance of all test methods with 4x upscaling is slightly decreased.
%is still decreased, but not much.
%In the first experiment, \textcolor{green}{As seen in Table~\ref{table3} our m-SRCNN model both Up-Bilinear and Up-Bicubic outperforming all common existing enlargement and original SRCNN at 2x upscaling. Especially our m-SRCNN Up-Bilinear model. For 3x upscaling, Original SRCNN Up-Bicubic model achieve the best performace. For 4x upscaling, the conventional upscaling method, bicubic, achieve the best performace.

\begin{figure}[htbp]
\centerline{\includegraphics[width=1\linewidth]{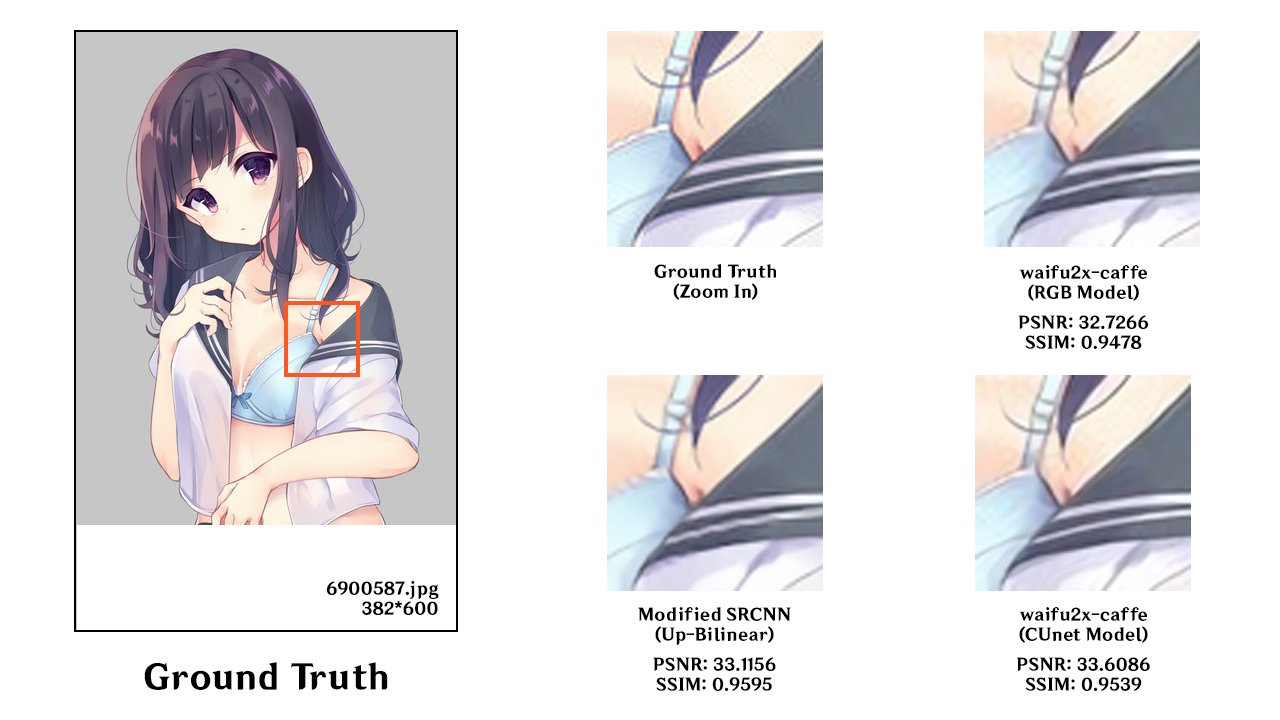}}
\caption{A comparison result of our m-SRCNN with up-bilinear and waifu2x models, when the test image is enlarged with 2x upscaling.}
\label{fig:fig10a}
\end{figure}

In the second experiment, our baseline is an online application, waifu2x. As can be seen in Table~\ref{table4}, our m-SRCNN with up-bilinear model outperforms the waifu2x in all upscaling factors and in all cases. Fig.~\ref{fig:fig10a} shows visual test results.

\begin{table*}[ht!]
\begin{adjustwidth}{-0.75cm}{-0.75cm}
\centering
\caption{Average PSNR and SSIM in comparison of different enlargement methods and different upscaling factors.} 
%Average PSNR and SSIM of all images in Testing set from different enlargement methods and upscaling factor

\begin{center}
\begin{tabular}{c|c c c c c c c c c c c c c c}

\hline
\multicolumn{1}{c|}{}                   & \multicolumn{2}{c}{}         & \multicolumn{2}{c}{}                                                             & \multicolumn{2}{c}{}                                                             & \multicolumn{2}{c}{}              & \multicolumn{2}{c}{}                                                             & \multicolumn{2}{c}{\textbf{Modified}}                                            & \multicolumn{2}{c}{\textbf{Modified}}                                            \\
\multicolumn{1}{c|}{}                   & \multicolumn{2}{c}{Nearest}  & \multicolumn{2}{c}{Bilinear}                                                     & \multicolumn{2}{c}{Bicubic}                                                      & \multicolumn{2}{c}{SRCNN}         & \multicolumn{2}{c}{SRCNN}                                                        & \multicolumn{2}{c}{\textbf{SRCNN}}                                               & \multicolumn{2}{c}{\textbf{SRCNN}}                                               \\
\multicolumn{1}{c|}{}                   & \multicolumn{2}{c}{Neighbor} & \multicolumn{2}{c}{}                                                             & \multicolumn{2}{c}{}                                                             & \multicolumn{2}{c}{(Up-Bilinear)} & \multicolumn{2}{c}{(Up-Bicubic)}                                                 & \multicolumn{2}{c}{\textbf{(Up-Bilinear)}}                                       & \multicolumn{2}{c}{\textbf{(Up-Bicubic)}}                                        \\
\multicolumn{1}{c|}{}                   & \multicolumn{2}{c}{}         & \multicolumn{2}{c}{}                                                             & \multicolumn{2}{c}{}                                                             & \multicolumn{2}{c}{}              & \multicolumn{2}{c}{}                                                             & \multicolumn{2}{c}{\textbf{(Proposed)}}                                          & \multicolumn{2}{c}{\textbf{(Proposed)}}                                          \\ \cline{2-15} 
\multicolumn{1}{c|}{\multirow{-5}{*}{}} & PSNR           & SSIM        & PSNR                                    & SSIM                                   & PSNR                                    & SSIM                                   & PSNR             & SSIM           & PSNR                                    & SSIM                                   & PSNR                                    & SSIM                                   & PSNR                                    & SSIM                                   \\ \hline
\multicolumn{1}{c|}{2x}                 & 27.2484        & 0.9079      & 27.7958                                 & 0.8996                                 & 29.1088                                 & 0.9209                                 & 31.0977          & 0.9372         & 29.3876                                 & 0.9184                                 & {\color[HTML]{00B050} \textbf{31.2819}} & {\color[HTML]{00B050} \textbf{0.9410}} & {\color[HTML]{0070C0} \textbf{29.4130}} & {\color[HTML]{0070C0} \textbf{0.9213}} \\
\multicolumn{1}{c|}{3x}                 & 22.7810        & 0.8037      & {\color[HTML]{00B050} \textbf{24.5525}} & {\color[HTML]{00B050} \textbf{0.8308}} & 24.4020                                 & 0.8287                                 & 23.6992          & 0.8235         & {\color[HTML]{0070C0} \textbf{24.6550}} & {\color[HTML]{0070C0} \textbf{0.8328}} & 23.8254                                 & 0.8268                                 & 24.5767                                 & 0.8320                                 \\
\multicolumn{1}{c|}{4x}                 & 22.3578        & 0.7744      & {\color[HTML]{00B050} \textbf{23.8173}} & 0.7926                                 & {\color[HTML]{0070C0} \textbf{23.9540}} & {\color[HTML]{0070C0} \textbf{0.7995}} & 23.7808          & 0.8006         & 23.8359                                 & 0.7941                                 & 23.7688                                 & {\color[HTML]{00B050} \textbf{0.8018}} & 23.7940                                 & 0.7939                                 \\ \hline
\multicolumn{15}{l}{\textcolor{ForestGreen}{Green} is the best result for the bilinear model, while \textcolor{NavyBlue}{Blue} is the best result for the bicubic model.} \\

\end{tabular}
\label{table3}
\end{center}
\end{adjustwidth}
\end{table*}

\begin{table*}[ht!]
\centering
\caption{Average PSNR and SSIM in comparison of m-SRCNN with up-bilinear, waifu2x (RGB), and waifu2x (CUnet) models in three different upscaling factors.}
%Average PSNR and SSIM of all images in testing set comparison between m-SRCNN (Up-Bilinear), waifu2x (RGB), and waifu2x (CUnet) model with different upscaling factors

\begin{center}
\begin{tabular}{c|cccccc}
\hline
          & \multicolumn{2}{c}{\textbf{Modified SRCNN}} & \multicolumn{2}{c}{waifu2x-caffe} & \multicolumn{2}{c}{waifu2x-caffe} \\
Upscaling & \multicolumn{2}{c}{\textbf{(Up-Bilinear)}}  & \multicolumn{2}{c}{(RGB Model)}   & \multicolumn{2}{c}{(CUnet Model)} \\
Factor    & \multicolumn{2}{c}{\textbf{(Proposed)}}     & \multicolumn{2}{c}{}              & \multicolumn{2}{c}{}              \\ \cline{2-7} 
          & PSNR                  & SSIM                & PSNR             & SSIM           & PSNR             & SSIM           \\ \hline
2x        & \textbf{31.0103}      & \textbf{0.9377}     & 29.1175          & 0.9032         & 29.5020          & 0.9068         \\
3x        & \textbf{23.6713}      & \textbf{0.8268}     & 22.7281          & 0.7942         & 22.9847          & 0.7988         \\
4x        & \textbf{23.5286}      & \textbf{0.7960}     & 22.2065          & 0.7689         & 22.5755          & 0.7766         \\ \hline
\end{tabular}
\label{table4}
\end{center}
\end{table*}

\begin{figure}[htbp]
\centerline{\includegraphics[width=1\linewidth]{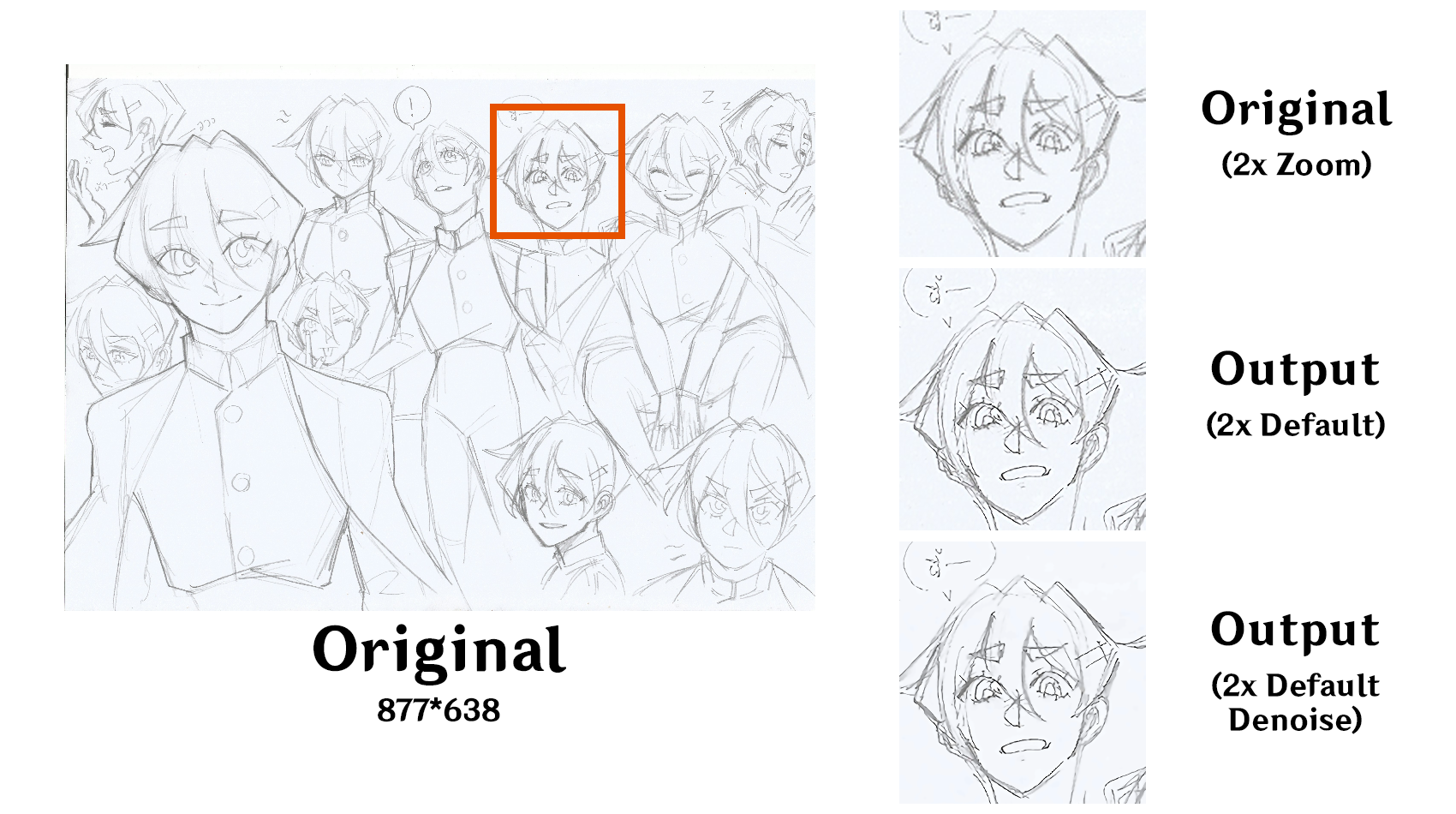}}
\caption{An example of original, output, and denoised images of an unseen illustration “Dijel” using m-SRCNN with up-bilinear model.}
\label{fig:fig7}
\end{figure}

\begin{figure}[htbp]
\centerline{\includegraphics[width=1.1\linewidth]{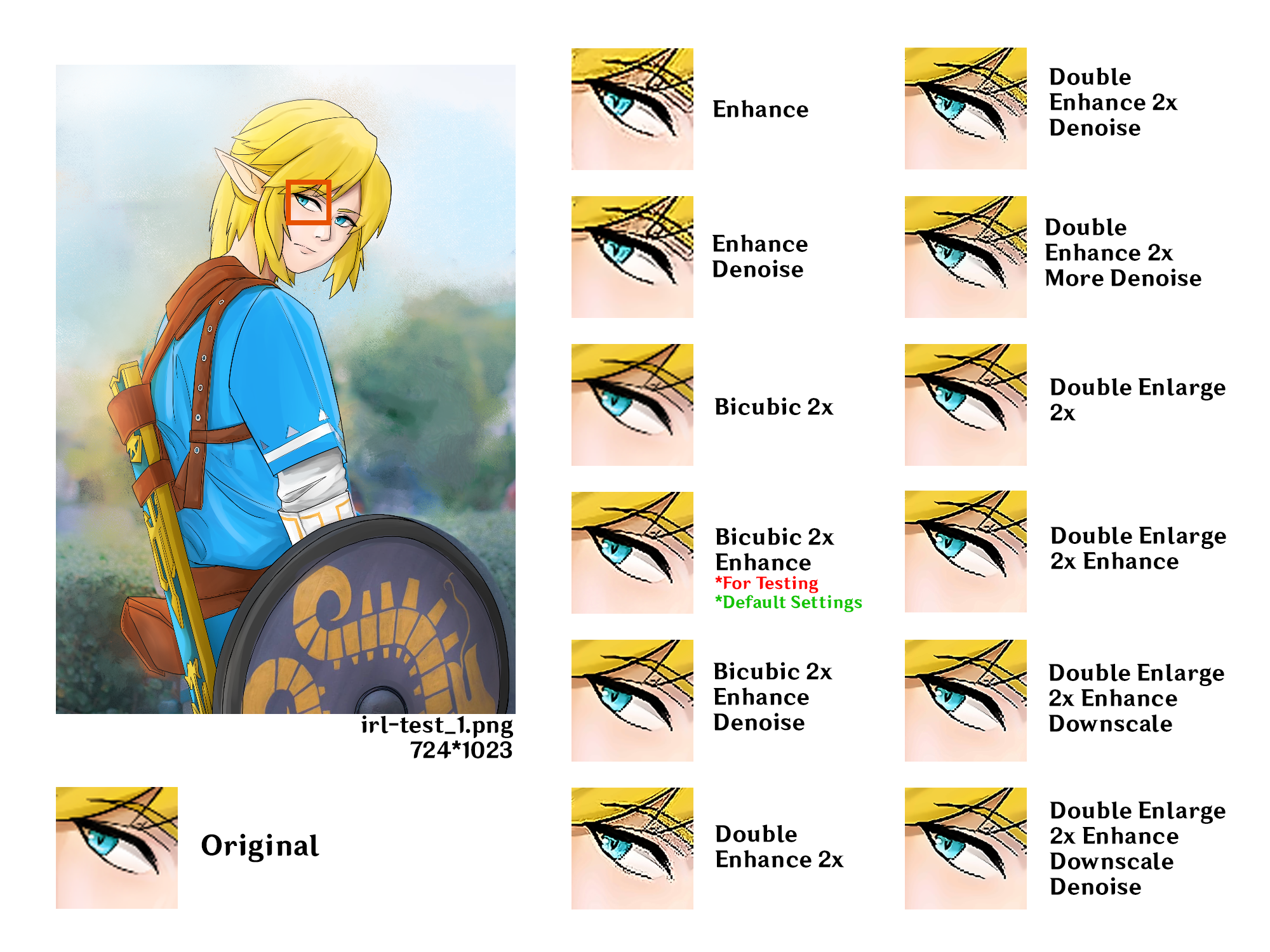}}
\caption{Results of four post-processing modules, including image-enhancement only, enlargement and enhancement, double enhancement, and double enlargement, when an unseen illustration “Link Cosplay” was tested by m-SRCNN with up-bilinear model.}
\label{fig:fig8}
\end{figure}

\begin{figure}[htbp]
\centerline{\includegraphics[width=1\linewidth]{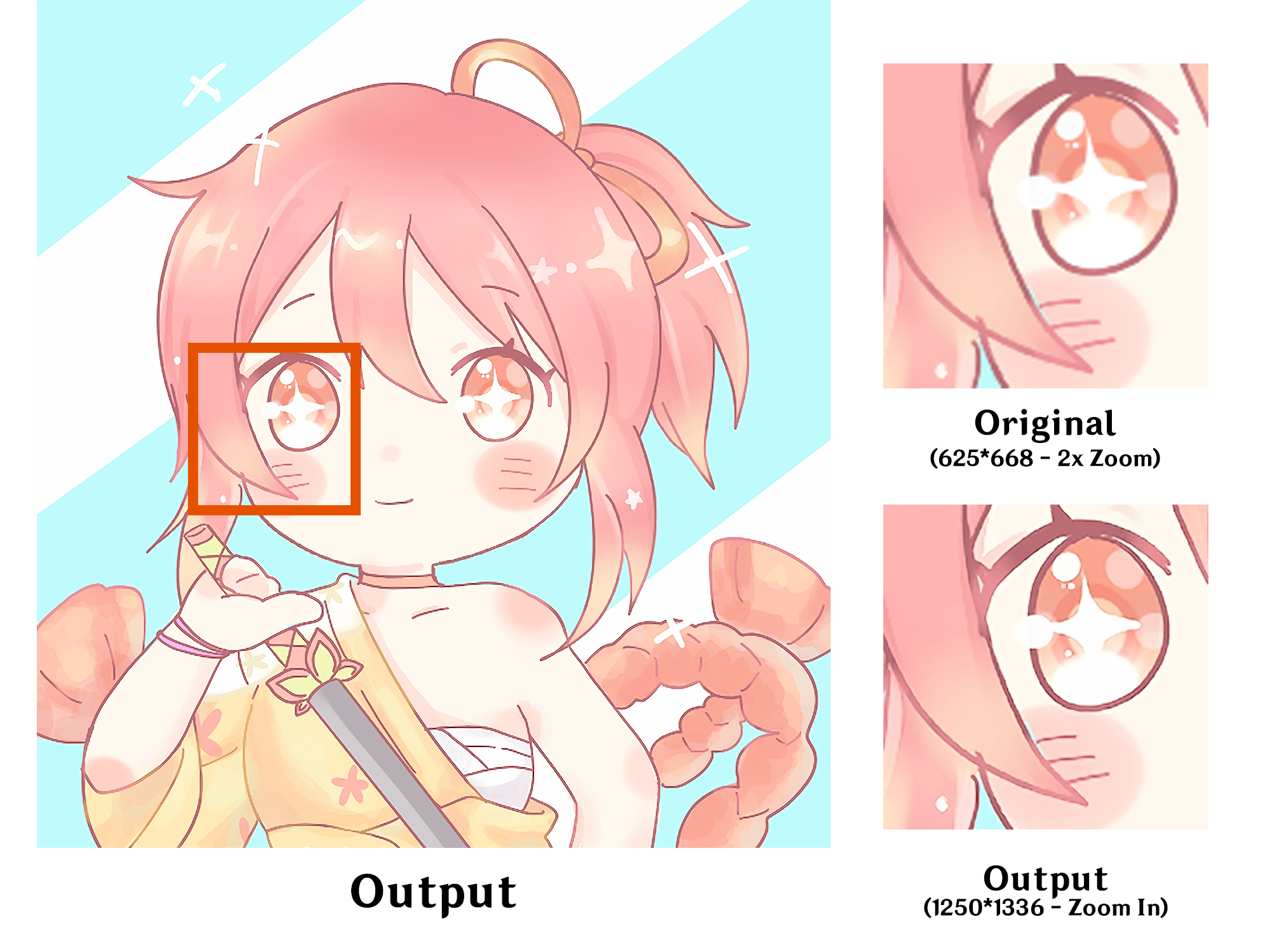}}
\caption{An example of enlarged unseen illustration “Miyagami-San” using m-SRCNN with default settings.}
\label{fig:fig9}
\end{figure}

In real-world use, users can input their images into a program without downscaling them. Hence, we must test it without a reference. The visual test results show in Figs.~\ref{fig:fig7}, ~\ref{fig:fig8}, and~\ref{fig:fig9}, when the test images were enlarged with 2x upscaling.

By comparing m-SRCNN with up-bilinear model and waifu2x, we found that our model performs better than the baseline, because of adding a step of image preprocessing (image sharpening and denoising) also with the different upscaling methods in the model training. Furthermore, all images were pre-upsampled \cite{DL-survey-1} before the model enhancement. 
%By comparing our Up-Bilinear and waifu2x models, we conclude that our models perform better because of image preprocessing in the model training step (Image sharpening and denoising) also with the different upscaling methods. For our models, the image has been pre-upsampled \cite{DL-survey-1} before model enhancement, this is different from waifu2x which image has been upsampled in the model. \cite{waifu2x-github} Image scaling artifacts \cite{iqa-srcnn} also found in waifu2x models.

\begin{figure*}[ht!]
\centerline{\includegraphics[width=1\linewidth]{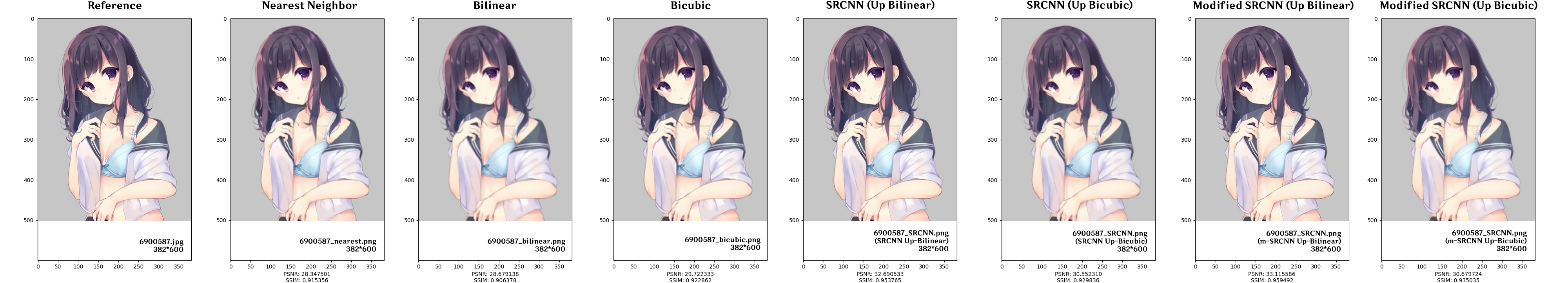}}
\caption{A visual comparison of result images of baseline and our methods with 2x upscaling, when compared with ground truth image (no enlargement).}
\label{fig:fig10}
\end{figure*}

\section{Conclusion}
This paper has proposed a modified SRCNN (m-SRCNN) with up-bilinear and up-bicubic models. The previous approach, like waifu2x using SRCNN model, for image enlargement suffered from a lack of ability to enhance the quality of enlarged images effectively. 
%\textcolor{red}{if the model do not enlarge the input image.} 
Hence, the m-SRCNN was designed to fulfill this gap. Responding to this, in our design, the key components, that can be viewed as the main contributions of this paper, are listed below: \footnote{Source code available: https://github.com/TanakitInt/SRCNN-anime}

\begin{itemize}
\item A denoised and sharpened dataset was added for model training.
\item Image enhancement-only function was created.
\item Double enhancement and double enlargement functions were developed.
\item User customization for image scaling, image denoising, double enhancement, and double enlargement was provided.
\end{itemize}

As shown in experimental results, our m-SRCNN can outperform the original SRCNN and waifu2x in most cases.

In future work, we aim to investigate another issue to 
%a web application for ease of use, also increase the number of datasets by adding images from a different \textcolor{green}{data source}. This should 
allow us to inspect the diversity of the Anime-style art images. Furthermore, we desire to upgrade our work to achieve a faster and more accurate model.
%super-resolution Convolutional Neural Network (FSRCNN) \cite{10.1007/978-3-319-46475-6_25} to decrease the operation time and Super-Resolution Generative Adversarial Network (SRGAN) \cite{jin2017towards} for better image quality.

\section*{Acknowledgment}
This work is affiliated with Image Processing and Deep Learning Laboratory (IPDL Lab) and Anime Cosplay and Boardgame Club (ABOARD Club), Faculty of Information Technology, King Mongkut's Institute of Technology Ladkrabang. Laboratory hardware and Anime images from ABOARD club were used during the development. Thanks for the original work Anime-style art images from Nikamon Saelim, Apinyarut Manakul, and Patharapan Hongtawee.

%section*{References}
\bibliographystyle{IEEEtran}
\bibliography{IEEEabrv, bibliography}

\end{document}